\documentclass[twocolumn,prb,aps,10pt,superscriptaddress]{revtex4-1}

\usepackage{graphicx}
\usepackage{amsmath}
\usepackage{amsfonts}
\usepackage{amssymb}
\usepackage{xcolor}
\usepackage{hyperref}
\hypersetup{colorlinks=true,allcolors=blue}
\usepackage[T1]{fontenc}

\usepackage{footnote}
\usepackage{textcomp}

\newcommand{\mbf}[1]{\mathbf{#1}}
\renewcommand{\t}[1]{\textrm{#1}}
\newcommand{\nn}{\nonumber\\}

\newcommand{\q}{\mbf{q}}

\renewcommand{\a}{\alpha}

\newcommand{\g}{\gamma}

\newcommand{\n}{\nu}

\renewcommand{\r}{\rho}
\newcommand{\s}{\sigma}

\newcommand{\w}{\omega}

\newcommand{\D}{\Delta}

\newcommand{\barr}{\bar{\r}}

\renewcommand{\L}{\mathcal{L}}

\newcommand{\+}{^\dagger}
\renewcommand{\>}{\rangle}

\newcommand{\Tr}{\t{Tr}}
\newcommand{\G}{\vert G\>}
\newcommand{\X}{\vert X\>}

\newcommand{\XX}{\vert X\>\<X\vert}
\newcommand{\XG}{\vert X\>\<G\vert}
\newcommand{\GX}{\vert G\>\<X\vert}

\begin{document}

\title{Transiently changing shape of the photon number distribution in a\\
  quantum-dot--cavity system
  driven by chirped laser pulses}

%Alternative title suggestions:

\author{M. Cosacchi}
%\email{michael.cosacchi@uni-bayreuth.de}
\affiliation{Theoretische Physik III, Universit{\"a}t Bayreuth, 95440 Bayreuth, Germany}
\author{T. Seidelmann}
\affiliation{Theoretische Physik III, Universit{\"a}t Bayreuth, 95440 Bayreuth, Germany}
\author{F. Ungar}
\affiliation{Theoretische Physik III, Universit{\"a}t Bayreuth, 95440 Bayreuth, Germany}
\author{M. Cygorek}
\affiliation{Department of Physics, University of Ottawa, Ottawa, Ontario, Canada K1N 6N5}
\author{A. Vagov}
\affiliation{Theoretische Physik III, Universit{\"a}t Bayreuth, 95440 Bayreuth, Germany}
\affiliation{ITMO University, St. Petersburg, 197101, Russia}
\author{V. M. Axt}
\affiliation{Theoretische Physik III, Universit{\"a}t Bayreuth, 95440 Bayreuth, Germany}

\begin{abstract}

We have simulated the time evolution of the photon number
distribution in a semiconductor quantum dot-microcavity system driven by chirped laser pulses
and compare with unchirped results.  When phonon interactions with the dot are
disregarded - thus corresponding to the limit of atomic cavity systems - chirped
pulses generate  photon number distributions that change their shape drastically
in the course of time.
Phonons have a
strong and qualitative impact on the photon statistics.
The asymmetry
between phonon absorption and emission destroys the symmetry of the photon
distributions obtained for positive and negative chirps.
While for negative chirps transient distributions resembling thermal ones are observed, for positive chirps the photon
number distribution still resembles its phonon-free counterpart but with overall
smoother shapes.
In sharp contrast, using unchirped pulses of the same pulse area and 
duration
wave-packets are found that move up and down the Jaynes-Cummings
ladder with a bell-shape that changes little in time.
For shorter pulses and lower driving strength Rabi-like oscillations
occur between low photon number states.
For all considered excitation conditions transitions
between sub- and super-Poissonian statistics are found at certain times.
For resonant  driving with low intensity the
Mandel parameter oscillates and is mostly negative, which indicates a non-classical state in the cavity field.
Finally, we show that it is possible that the Mandel parameter 
dynamically approaches
zero and still the photon distribution exhibits two maxima and thus is
far from being a Poissonian.
\end{abstract}

\maketitle

\section{Introduction}
\label{sec:Introduction}

Semiconductor quantum-dot--cavity (QDC) systems continue to raise attention as
highly integrable on-demand emitters of non-classical states of light.  In
particular, QDCs have proven to be rather successful providing, e.g., reliable
on-demand high quality single photon sources
\cite{Michler2000,Santori2001,Santori2002,He2013,Wei2014Det,Ding2016,Somaschi2016,Schweickert2018,Hanschke2018,Cosacchi2019}
as well as sources for entangled photon pairs
\cite{Akopian2006,Stevenson2006,Hafenbrak2007,Dousse2010,delvalle2013dis,Mueller2014,Orieux2017,Seidelmann2019}.
Clearly, QDCs support a much larger class of excitations when higher mean photon
numbers are reached.  The additional degrees of freedom provided by higher
number photon states obviously allow for a rich variety of dynamical scenarios
and may open the way to new kinds of applications such as, e.g., the encoding of
quantum information in the photon number state distribution. These possibilities
are, however, far from being explored.

Often, the first step to characterize systems with photon distributions ranging
up to higher photon numbers is to record a few characteristic numbers such as
the mean photon number \cite{Chough1996} and/or the Mandel parameter
\cite{Mandel1979}.  In simple cases, the mean photon number is indeed enough to
capture the whole information about the photon distribution even when the latter
is time dependent. This applies in particular when photons are generated by
classically driving an empty cavity without a quantum dot (QD) where the photonic system is at all times in a coherent
state and thus the distribution is a Poissonian \cite{Gerry2004}, i.e., in this
case the photonic excitation is always as close as possible to a classical light
field and thus non-classical states cannot be reached. Moreover, although the
mean photon number varies in time, the photon distribution keeps its shape at
all times.

The situation is different when a system with few discrete levels near resonance
to a cavity mode such as an atom or a quantum dot is placed inside the
cavity. When driving transitions between these discrete levels deviations from
the coherent state may occur as is evident, e.g., by monitoring the Mandel parameter
\begin{align}
  Q(t)=(\langle \D n^2\rangle-\langle n\rangle)/\langle n\rangle\, .
\end{align}
$Q(t)$ measures the deviation of the mean-square fluctuation
from the mean photon number normalized to the latter.
Therefore, $Q$ vanishes for a Poisson distribution.
A positive $Q$ indicates a super-Poissonian distribution with larger
fluctuations than in a coherent state with the same mean photon number
while negative $Q$ values correspond to
the sub-Poissonian regime which is known to have no classical analog
\cite{Vogel2006}.  Indeed, deviations from the coherent state have been reported
for the stationary distribution obtained in an atomic cavity with constant driving
where different signs of $Q$ have been found for different ratios between cavity
loss and radiative decay rates \cite{Quang1993}.  In
Ref.~\onlinecite{Callsen2013} it has been shown that the statistics of photons
emitted from the exciton-biexciton system of a QD can be steered from
sub- to the super-Poissonian by varying the biexciton binding energy, the pump
strength or the temperature.
\footnote{Although the experiments in
Ref.~\onlinecite{Callsen2013} have been performed on  QDs without
cavity, the number of modes in the theoretical modeling was restricted to two
which corresponds to the situation in a QDC.  Therefore, the results should also
apply to QDCs.}
Simulations for a pulsed excitation of a QDC indicate that $Q$
can exhibit oscillations and change its sign repeatedly in time
\cite{Harouni2009}.

It is clear, however, that in general  the photon number distribution contains
much more detailed information than captured by the mean photon number or the
Mandel parameter.  Recently, calculations of the stationary photon number
distribution in a constantly driven QDC revealed a strong qualitative influence
of phonons on the shape of the distribution \cite{Cygorek2017,Reiter2019}.
While without phonons distributions with many different shapes were found for
different detunings, the stationary distribution with phonons turned out to be
close to a thermal state with a high effective temperature.

Advances in
measuring techniques have demonstrated possibilities for observing directly the
photon number resolved distributions in various systems, ranging from bimodal
microlasers \cite{Schlottmann2018} over QDs \cite{Schmidt2018,Helversen2019} to
exciton-polariton condensates \cite{Klaas2018}.  Furthermore, a novel algorithm
for data evaluation free of systematic errors to obtain number distributions has
been successfully employed \cite{Hlousek2019}.  These achievements could pave
the way to access the information encoded in photon number distributions with
unprecedented detail.

The focus of the present paper is on the transient behavior of the photon number
distribution in a QDC system driven by chirped pulses in comparison to the
unchirped case. Our most striking result is the finding that the shape of the
number distribution changes dynamically when driving the QDC with chirped
pulses.  In sharp contrast, for sufficiently strong unchriped excitations a
wave-packet which keeps a bell shape for all times moves up and down the
Jaynes-Cummings ladder. Phonons have noticeable effects on the photon statistics
for all excitation conditions that we compare. Notably, for chirped excitation
the phonon impact induces qualitative changes of the shape of the distribution
in particular for negative chirps.

\section{Theory}

\subsection{Model and Methods}
\label{subsec:Model}

We study a self-assembled QD, e.g., GaAs/In(Ga)As, with strong electronic
confinement, such that only the lowest conduction and the highest valence band
states need to be taken into account.  Furthermore, we consider only situations
where the system is well represented by a two-level model. The latter applies,
e.g., for resonant driving of the exciton by circularly polarized light when the
fine-structure splitting is negligible or when all other states such as the
biexciton are sufficiently far from resonance. Then the Hamiltonian for the
laser driven dot reads:
\begin{align}
\label{eq:H_DL_rotating}
  H_{\t{DL}}=&-\hbar\D\w_{\t{LX}}\XX \notag \\
            &-\frac{\hbar}{2}f(t)\left(e^{-i\varphi(t)}\XG+e^{i\varphi(t)}\GX\right)\, ,
\end{align}
where the detuning between the exciton and central laser frequency
$\D\w_{\t{LX}}:=\w_{\t{L}}-\w_{\t{X}}$ is introduced.  Here, the ground state
$\G$ is chosen as the zero of the energy scale.  Note that the usual dipole and
rotating wave approximations are employed and the Hamiltonian is written down in
a frame co-rotating with the laser frequency $\w_{\t{L}}$.  The real amplitude
$f(t)$ and the phase $\varphi(t)$ are related to the instantaneous Rabi frequency
$\Omega(t)$ by:
\begin{align}
  \Omega(t) := 2\mathbf{M_{0}}\cdot\mathbf{E}(t) = f(t)\,e^{-i(\omega_{\t{L}}t+\varphi(t))},
\label{Rabi-freq}
\end{align}
where $\mathbf{M_{0}}$ is the dipole matrix element of the
transition between the QD ground $\G$ and exciton state $\X$ and $\mathbf{E}$ is
the positive frequency part of the laser field.

To enhance the coupling between the QD and the electromagnetic field, the dot
can be placed into a microcavity.  We account for  a single cavity mode with
frequency $\w_{\t{C}}$ far from the electromagnetic continuum and a QD coupled
to that mode close to resonance via:
\begin{align}
\label{eq:H_C_rotating}
H_{\t{C}}=\hbar\D\w_{\t{CL}}a\+ a+\hbar g\left(a\+ \GX+a\XG\right)\, ,
\end{align}
where the cavity photons are created (annihilated) by the bosonic operator $a\+$ ($a$)
and are detuned by $\D\w_{\t{CL}}:=\w_{\t{C}}-\w_{\t{L}}$ from the laser frequency.
The QD is coupled to the cavity with a strength of $\hbar g$.

The subsystem of interest comprised of the dot-laser and the cavity Hamiltonian
$H_{\t{DL}}$ and $H_{\t{C}}$, respectively, is not an ideal few-level system,
since it is embedded into the surrounding solid state matrix.  Even at cryogenic
temperatures of a few Kelvin, the QD exciton is prone to the coupling to
phonons.  In strongly confined excitonic systems, the most important phononic
contribution usually results from the deformation potential coupling to longitudinal acoustic
(LA) phonons and is of the elastic pure dephasing-type
\cite{Besombes2001,Borri2001,Krummheuer2002,Axt2005}
\begin{align}
\label{eq:H_Ph}
H_{\t{Ph}}=\hbar\sum_\q \w_\q b_\q\+ b_\q+\hbar\sum_\q \left(\g_\q^{\t{X}}b_\q\+ +\g_\q^{\t{X}*}b_\q\right)\XX\, ,
\end{align}
where the bosonic operator $b_\q\+$ ($b_\q$) creates (destroys)
phonons with frequency $\w_\q$.  $\g_\q^{\t{X}}$ denotes the coupling constant
between the exciton state and the bosonic mode labeled by its wave vector $\q$
which is adequate for bulk phonons.  Here, we use the fact that in GaAs/In(Ga)As
the lattice properties of the dot and its surroundings are similar, such that
phonon confinement is negligible. Other QD-phonon interaction mechanisms like,
e.g., the piezo-electric coupling to LA and transverse acoustic (TA) phonons can become important in
strongly polar crystals such as, e.g., GaN-based QDs
\cite{Ostapenko2012,Krummheuer2005}, but are of minor importance  for GaAs-type
structures.

Finally, we account for Markovian loss processes by phenomenological decay rates
for the radiative decay and cavity losses, respectively, that are incorporated
into the model as Lindblad-type super-operators
$\L_{\GX,\g}\bullet+\L_{a,\kappa}\bullet$ with
\begin{align}
\mathcal{L}_{O,\Gamma}\bullet=\Gamma\left(O\bullet O\+ -\frac{1}{2}\left\lbrace\bullet,O\+ O\right\rbrace_+\right)\, ,
\end{align}
where $\{\cdot,\cdot\}_+$ denotes the anti-commutator.  $O$ is a
system operator and $\Gamma$ the decay rate of the associated loss process,
i.e., in our case $\gamma$ stands for the radiative decay rate while $\kappa$ is
the cavity loss rate.

The dynamical equation to be solved is the Liouville-von Neumann equation for
the density matrix
\begin{align}
  \label{eq:Liouville-von Neumann}
  \frac{\partial}{\partial t} \r =
-\frac{i}{\hbar}\{H,\r\}_- + \L_{\GX,\g}\r + \L_{a,\kappa}\r
\end{align}
with the total Hamiltonian $H=H_{\t{DL}}+H_{\t{C}}+H_{\t{Ph}}$ and
$\{\cdot,\cdot\}_-$ denotes the commutator.

We employ a path-integral formalism for simulating the dynamics in
the above defined model in a numerically complete fashion.  By tracing out the
phonon degrees of freedom analytically, a non-Markovian memory kernel decaying
on a time scale of a few picoseconds is obtained that manifests in experiments
as, e.g., non-Lorentzian line shapes in linear and non-linear
spectra\cite{Borri2001,Krummheuer2002,Vagov2004,Reiter2017} or in characteristic
dependencies of the phonon-induced damping of Rabi-rotations
\cite{Foerstner2003,Kruegel2005,Vagov2007,Ramsay2010a}.  Therefore, this memory
cannot be neglected in calculating the QD dynamics which takes place on a
similar time scale.  We call a numerical solution complete if a finer time
discretization or a longer cut-off of the phonon-induced memory kernel does not
change the results noticeably.

Most current implementations of the real-time path-integral approach are based
on the pioneering work of Makri and Makarov \cite{Makri1995a,Makri1995b}, who
introduced an iterative scheme for the augmented density matrix of the subsystem
of interest.  We are using an extension of this scheme that allows the inclusion
of non-Hamiltonian Lindblad-type contributions into the path-integral algorithm
without the loss of precision with respect to the phonon-induced part of the
dynamics by formulating the iterative scheme not in a Hilbert, but a Liouville
space \cite{Barth2016}.  In the present study, the system that couples to the
phonons is represented by a large number of basis states of the form
$|G,n\rangle$ and $|X,n\rangle$ where $n$ denotes the photon number and $G$ or
$X$ indicates whether the dot is in its ground or excited state.  A numerically
complete study of such systems is currently impossible with the Makri-Makarov
algorithm due to the extreme growth of the numerical demand with rising number
of system states.  Nevertheless, we are able to present numerically complete
results because we are using a recently developed reformulation of the algorithm
that iterates a partially summed augmented density matrix \cite{Cygorek2017}.
Note that this reformulation of the path-integral algorithm does not introduce
any additional approximations.  For details on the methods, consider the
supplement of Ref.~\onlinecite{Cygorek2017}.  The photon number distribution is
obtained by taking the corresponding matrix element of the subsystem's reduced
density operator $\barr=\Tr_{\t{Ph}}[\r]$, with $\Tr_{\t{Ph}}$ denoting the
trace over the phonon degrees of freedom,
\begin{align}
P_n(t) = \sum_{\n=\t{G,X}} \langle\nu,n|\barr(t)|\nu,n\rangle.
\end{align}

\subsection{Chirped pulses and laser-dressed states}
\label{subsec:Distributions}

In order to generate a chirped pulse one usually starts with a Gaussian pulse
with an envelope and phase:
\begin{align}
  \label{f-gauss}
  f_{0}(t)=\,& \frac{\Theta}{\sqrt{2\pi}\s} e^{-\frac{(t-t_{0})^2}{2\s^2}}\\
  \varphi(t)=\,&\t{const.}\,,
\end{align}
where  $\Theta$ denotes the pulse area and $\s$ determines the
duration corresponding to a full width at half maximum (FWHM) of
FWHM$=2\sqrt{2\ln(2)}\s$ and $t_{0}$ marks the time of the pulse maximum.  We
shall assume in the following a resonant excitation where $\varphi(t)=0$ in
Eq.~(\ref{Rabi-freq}) for an unchirped pulse.  We note in passing that also
other pulse shapes are possible as starting point for the generation of chirped
pulses. In particular, secant hyperbolic pulses may have advantages in certain
circumstances \cite{Melinger1995}.

Passing the initial pulse in Eq.~(\ref{f-gauss}) through a Gaussian chirp
filter\cite{Saleh2019} yields a chirped pulse with envelope and phase:
\begin{align}
\label{eq:chirped_function_Lab}
  f_{\t{chirp}}(t)=\,&\frac{\Theta_{\t{chirp}}}{\sqrt{2\pi}\s_{\t{chirp}}}
  e^{-\frac{(t-t_{0})^2}{2\s_{\t{chirp}}^2}}\\
\label{phi-chirp}
    \varphi(t) =\,& a\,(t-t_{0})^{2}/2\,,
\end{align}
pulse area $\Theta_{\t{chirp}}=\Theta\sqrt{\s_{\t{chirp}}/\s}$ and
duration $\s_{\t{chirp}}=\sqrt{(\a^2/\s^2)+\s^2}$. The phase in
Eq.~(\ref{Rabi-freq}) has acquired a quadratic time dependence,
which corresponds to an instantaneous laser frequency
$\w_{\t{L}}+\dot{\varphi}=\w_{\t{L}}+a\,(t-t_{0})$ that changes linearly in time
and for $\w_{\t{L}}=\w_{\t{X}}$ crosses the exciton resonance at the pulse
maximum $t=t_{0}$.  The strength of the chirp is commonly expressed in terms of
the chirp parameter $\a$ which is related to the coefficient $a$ in
Eq.~(\ref{phi-chirp}) by $a=\a/(\a^2+\s^4)$. Note that the pulse area and in
particular the pulse length increases drastically when chirps are introduced
(cf.~the definition of $\s_{\t{chirp}}$).

\section{Numerical Results on Transient photon statistics}
\label{sec:Results}

\begin{figure*}[t]
	\centering
	\includegraphics[width=0.9\textwidth]{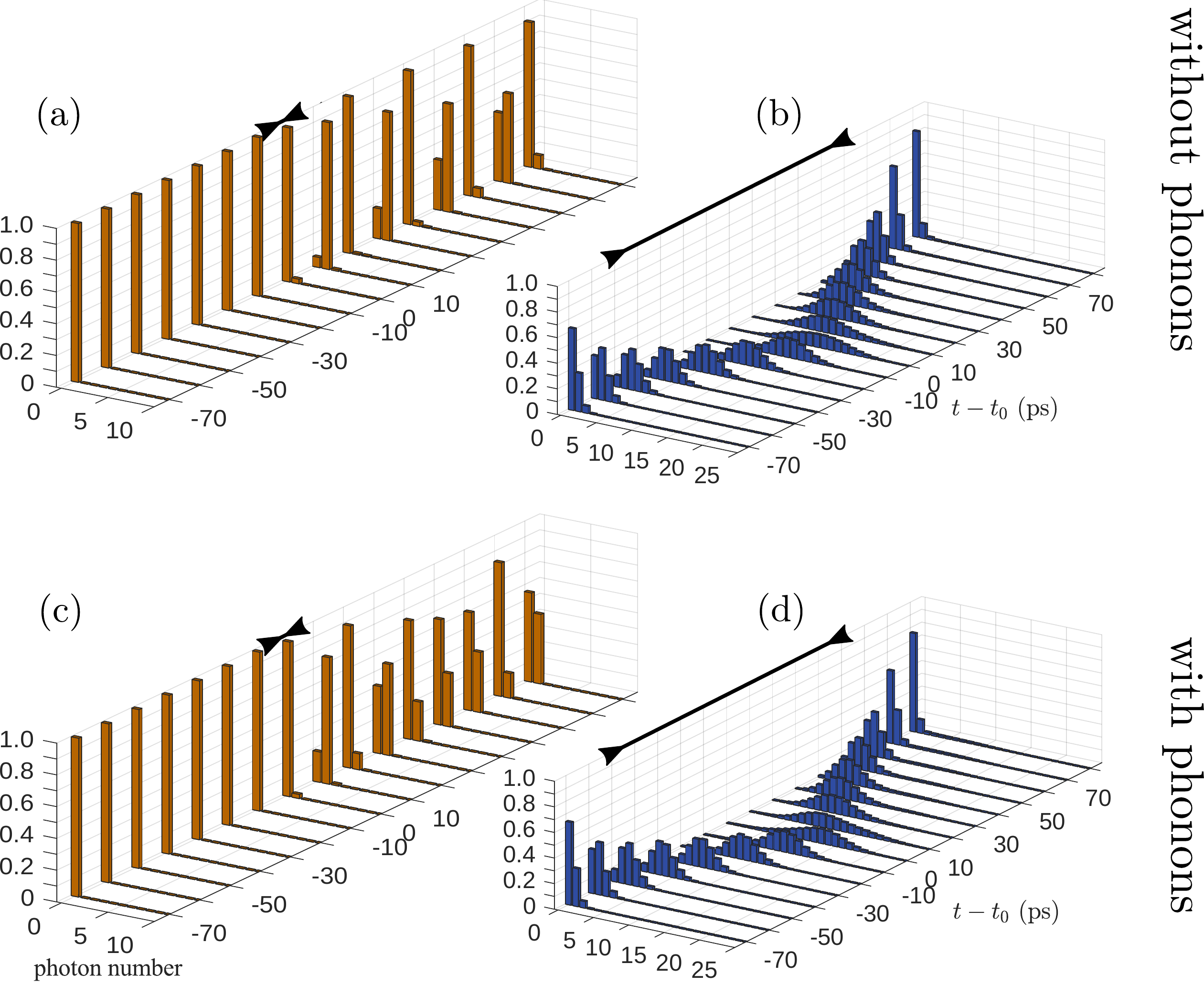}
	\caption{Transient photon number distributions for laser excitations
with unchirped pulses with: (a) and (c) pulse area $\Theta=5\pi$ and duration
FWHM=2.4 ps, (b) and (d) pulse area $\Theta=31.63\pi$ and duration FWHM$=$94.22 ps.
%In panels (b) and (d) the pulse area and FWHM have been chosen equal to
%the values one would obtain after passing the pulse used in (a) and (c) through a
%Gaussian chirp filter with chirp parameter $|\alpha|=40$ resulting in
%$\Theta\approx 31\pi$ and duration FWHM$\approx$92.5 ps. 
Panels (c) and (d) display
results accounting for phonons that are initially at equilibrium at a
temperature of $T=4\,$K while the corresponding phonon-free results are shown in
(a) and (b). The pulse has its maximum at $t=t_{0}$. Black markers indicate the FWHM
of the pulse.}
	\label{fig:chirp0}
\end{figure*}

For the numerical calculations, we assume a QD with $6\,$nm diameter and
standard GaAs parameters \cite{Krummheuer2005,Cygorek2017}.  The cavity is
coupled to the QD exciton with a strength of $\hbar g=0.1\,$meV while it is on
resonance, i.e., $\Delta\w_{\t{CX}}:=\w_{\t{C}}-\w_{\t{X}}=0$.  The cavity
losses are taken to be $\hbar\kappa=6.6\,\mu$eV, which corresponds to a quality
factor $Q\approx 10^5$ assuming a mode frequency of $\hbar\w_{\t{C}}=1.5\,$eV.  The
radiative decay rate of the QD exciton is set to $\hbar\g=2\,\mu$eV.

\subsection{The chirp-free situation}

Let us first concentrate on the chirp-free case.  Figure \ref{fig:chirp0} (a) and (c)
display photon number distributions at different times for a QDC driven by an
unchirped Gaussian pulse with a pulse area of $5\pi$ and a duration of $2.4\,$ps
FWHM.  Panel (a) shows results without phonons while in panel (c) the
corresponding simulations with phonons are depicted assuming the phonons before
the pulse to be in thermal equilibrium  at a temperature of T$=4\,$K. The
initial state for the cavity photons is taken to be the vacuum, i.e., the $n=0$
Fock state and the QD is initially in the ground state.

\begin{figure*}[t]
	\centering
	\includegraphics[width=0.9\textwidth]{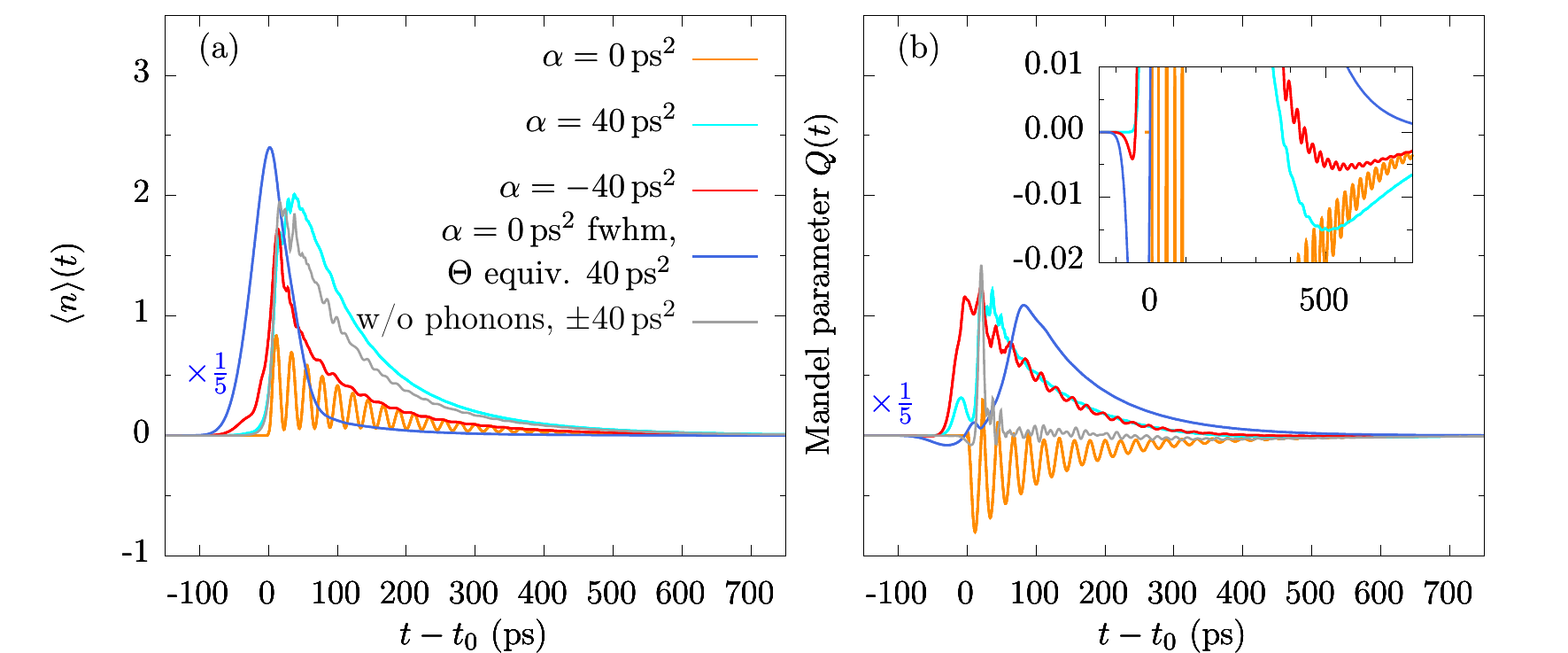}
	\caption{The time dependent (a) mean photon number and (b)  Mandel
parameter $Q(t)=(\langle \D n^2\rangle-\langle n\rangle)/\langle n\rangle$ for
the cases indicated by the labels.  All curves are calculated with phonons initially
at $T=4$ K, except for the grey curves which correspond to the phonon-free case.
The blue curve is scaled down by a factor of 5 for better visibility.
The inset in (b) corresponds to a zoomed-in scale.}
	\label{fig:n_Q_fig_dyn}
\end{figure*}

As expected the photons stay in the vacuum state until the arrival of the pulse.
At the end of the $5\pi$ pulse (cf.~black markers in Fig.~\ref{fig:chirp0}) the
QD is in the exciton state and the resonant coupling to the cavity initiates
vacuum Rabi oscillations
\cite{reithmaier2004str,yoshie2004vac,khitrova2006vac,Nahri2017,Kuruma2018},
i.e., oscillations between the $|X,n=0\rangle$ and the $|G,n=1\rangle$ states.
This is reflected in the photon distribution as oscillations between the $n=0$
and $n=1$ Fock states and results in damped oscillations of the mean photon
number  between zero and and a maximal amplitude that due to losses and phonon
effects is below one [cf.~orange curve in Fig.~\ref{fig:n_Q_fig_dyn} (a)].
Quantitatively, a small occupation of the two-photon state $|2\rangle$ is
observed, seen e.g., for $t-t_0=10\,$ps in Fig.~\ref{fig:chirp0} (a) and (c).
The reason lies in the re-excitation of the QD during the same pulse, whereby
effectively two photons can be put into the single cavity mode.

The phonon impact on Rabi-type oscillations in a two-level system has been
extensively studied \cite{Foerstner2003,Machnikowski2004,Kruegel2005,Kruegel2006,Vagov2007,mogilevtsev2008dri,mogilevtsev2009non,Ramsay2010a,Ramsay2010b,mccutcheon2010qua,mccutcheon2011gen,Reiter2019} and shall therefore not be analyzed here in detail.
We just note that the main effects are a phonon-induced damping, which depends on the driving strength, and a
renormalization of the Rabi frequency. The renormalization of $g$ is reflected
in Fig.~\ref{fig:chirp0} (a) and (c) by slightly different oscillation frequencies.
The damping seen in the orange curve in Fig.~\ref{fig:n_Q_fig_dyn} (a)
is the result of the combined effects of phonons, cavity losses and radiative decay.

\begin{figure*}[t]
	\centering
	\includegraphics[width=0.9\textwidth]{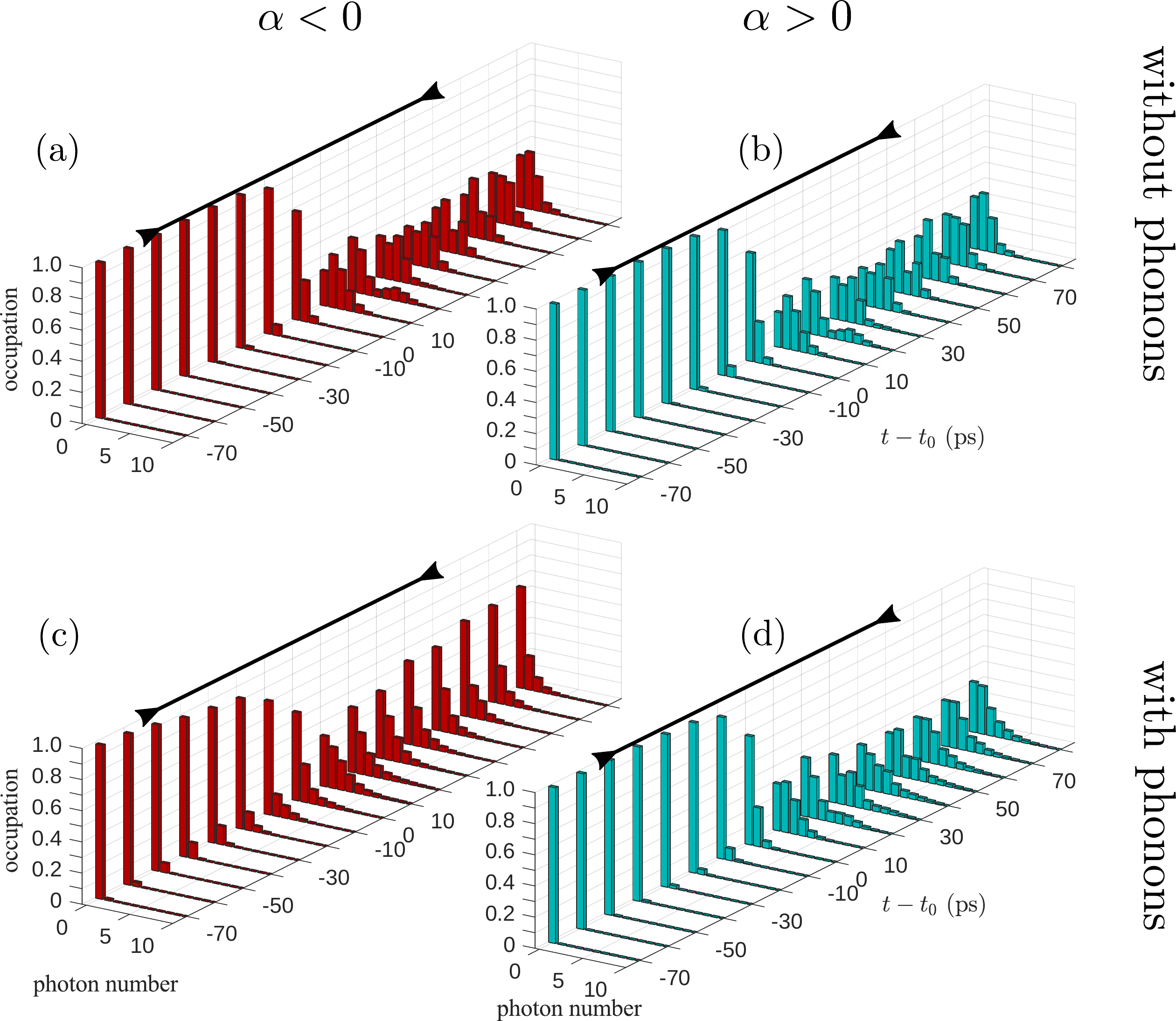}
	\caption{Transient photon number distributions for laser excitations
with chirped pulses with pulse area and FWHM before the chirp filter of:
$\Theta=5\pi$ and  FWHM=2.4 ps, i.e., $\Theta_{\t{chirp}}=31.63\pi$ and duration FWHM$_{\t{chirp}}=$94.22 ps for $|\alpha|=40$ ps$^2$.  (a) and (c) calculated with chirp parameter
$\alpha=-40$ ps$^2$, (b) and (d) $\alpha=+40$ ps$^2$.  Panels (c) and (d) display results accounting for
phonons that are initially at equilibrium at a temperature of $T=4\,$K while the
corresponding phonon-free results are shown in (a) and (b). The pulse has its maximum
at $t=t_{0}$. Black markers indicate the FWHM of the pulse after the chirp filter.}
	\label{fig:chirp}
\end{figure*}

For a fair comparison between unchirped and chirped pulses,
recall that the application of a Gaussian chirp filter involves
besides the time-dependent variation of the phase $\varphi(t)$ in Eq.~\eqref{phi-chirp} also a
considerable increase of the pulse duration and of the pulse area.
Therefore, we show in Fig.~\ref{fig:chirp0} (b) and (d) the photon distribution with and
without the influence of phonons for a pulse with pulse area $\Theta=31.63\pi$ and duration FWHM$=$94.22 ps, which corresponds to the application of a filter with an
effective value of $|\alpha| = 40$ ps$^2$ in Eq.~\eqref{eq:chirped_function_Lab} but keeping the phase
$\varphi(t)=0$ constant.
%Panel (b) and (d) of Fig.~\ref{fig:chirp0} also refer to an unchriped excitation,
%but now with a pulse area $\Theta=31.63\pi$ and duration FWHM$=$94.22 ps that according to
%Eq.~(\ref{eq:chirped_function_Lab}) correspond to the effective values if $|\alpha|=40$ ps$^2$,
%i.e., the driving is now much stronger and lasts considerably longer.
Most strikingly, with this
driving there are no traces of vacuum Rabi oscillations visible. Instead, a
wave-packet-type dynamics sets in, where a bell-shaped distribution is found for
all times. The mean photon number rises monotonically in time to values
$n\approx12$ (note that the blue curve in Fig.~\ref{fig:n_Q_fig_dyn} (a) is
scaled down by a factor of 5 for better visibility) and subsequently falls back
to zero after the pulse has vanished.

\subsection{Finite chirps}
%We consider this case in order to be able to compare two excitation conditions where the only difference is the frequency modulation, while the pulse area and length are kept the same.

Figure \ref{fig:chirp} displays transient photon number distributions obtained
for chirped pulses that are generated by passing the Gaussian pulse used in Figs.~\ref{fig:chirp0} (a) and (c) through a chirp filter with $\alpha=\pm40$ ps$^2$ [(a) and (c) $\alpha=-40$ ps$^2$, (b) and (d) $\alpha=+40$ ps$^2$]. The upper
panels correspond to simulations without phonons while for the lower panels the interaction with phonons has been included.
Note that the pulses used in Fig.~\ref{fig:chirp} have the same pulse area and duration as the unchirped pulses used in Figs.~\ref{fig:chirp0} (b) and (d) which allows us to compare excitation conditions where the only difference is the frequency modulation.

In the phonon-free case identical distributions are
obtained for positive and negative chirp [cf.~Fig.~\ref{fig:chirp} (a) and (b)].
This symmetry is removed when phonons are taken into account
[cf.~Fig.~\ref{fig:chirp} (c) and (d)].  In contrast to the unchirped case with the
same pulse area and duration in Fig.~\ref{fig:chirp0} (b) and (d), the photon number
is close to zero until the pulse maximum is reached, which can be explained by noting
that for chirped pulses the instantaneous laser frequency is strongly detuned from the
QD resonance for times away from the pulse maximum. The most striking difference
compared with Fig.~\ref{fig:chirp0} (b) and (d) is, however, that the photon
distributions in Fig.~\ref{fig:chirp} significantly change their shape in time.
The distributions found in the phonon free-case have at early times after the
pulse maximum a bell-shape with a single maximum and transform into a bimodal
distribution with two well separated bell-shaped contributions at later times
[cf.~$t-t_{0}=20$ ps in Fig.~\ref{fig:chirp} (a)]. Subsequently, at times
$t-t_{0}\approx30\!-\!50$ ps the distribution still has two peaks but
looks rather jagged having little resemblance with bell-shaped distributions.
Eventually, at
later times only a single maximum is found which appears at a finite photon
number or at zero, depending on time.

Phonons change the situation qualitatively for negative chirp
[cf.~Fig.~\ref{fig:chirp} (c)], where now the photon number distribution has a
single maximum at $n=0$ for all times.  The shape of the transient distribution
resembles thermal photon occupations, which due to  mean photon numbers around
$n=2$ [cf.~Fig.~\ref{fig:n_Q_fig_dyn} (a)] corresponds to an effective temperature above
$T_{\t{eff}}\approx40\,000$ K for photon energies
$\hbar\omega_{\t{C}}\approx1.5$ eV .  A similar impact of phonons on the photon number
distribution has been reported previously for the stationary distribution found
at long times for permanent driving \cite{Cygorek2017}.  The phonon impact for
positive chirp is less dramatic [cf.~Fig.~\ref{fig:chirp} (d)].  As in the
phonon-free case, there are still times where the distribution is bi-modal while
at other times only a single maximum is found. Overall, the irregular looking
shape appearing at certain times in Fig.~\ref{fig:chirp} (a) and (b) is
smoothened. Moreover, there is a tendency to build up a maximum near $n=0$.

Further differences between the number distributions in Figs.~\ref{fig:chirp0}
and \ref{fig:chirp} are revealed by looking at the time evolution of the
corresponding Mandel parameters $Q(t)$ in Fig.~\ref{fig:n_Q_fig_dyn} (b).  For a Fock
state the number fluctuation disappears, leading to a negative Mandel parameter,
except for the $n=0$ Fock state, where the Mandel parameter approaches an
expression of the form zero divided by zero.  We see from the orange curves in
Fig.~\ref{fig:n_Q_fig_dyn} that for weakly driven unchirped pulses the damped
oscillation of the mean photon number between 0 and at most 1 is accompanied by
damped oscillations of the Mandel parameter ranging down to almost -1 and up to
essentially 0.
The negative values of the minima correspond
to times where the system is close to the $n=1$ Fock state.
If the dynamics would exclusively involve states with photon numbers $0$ or $1$
such that only $P_{0}$ and $P_{1}$ are different from zero, it is easy to show that
for all times, where $P_{1}\neq0$, the Mandel parameter is $Q(t)=-\langle n\rangle$.
Therefore, $Q$ should approach $0$ when the $n=0$ Fock state is approached.
We see, however, from the orange curve in Fig.~\ref{fig:n_Q_fig_dyn} (b)
that the first maxima of the Mandel parameter $Q$ are a bit above 0, indicating 
small admixtures of higher number states.

For higher pulse areas $Q$ is positive for most of the time for chirped as well
as for unchirped pulses.  Interestingly, although the bell-shaped distributions
in Fig.~\ref{fig:chirp0} (b) and (d) at first glance resemble much more
Poissonian distributions than the somehow irregular ones found for chirped
pulses in Fig.~\ref{fig:chirp} (b) and (d) their deviation from a Poissonian as
measured by the Mandel parameter is much larger than for chirped pulses (note
that the blue curve in Fig.~\ref{fig:n_Q_fig_dyn} (b) is scaled down by a factor
of 5).  But most remarkably, in the calculation with finite chirp without
phonons [cf.~the grey line in Fig.~\ref{fig:n_Q_fig_dyn} (b)] the Mandel
parameter decays extremely fast after its initial rise to positive values
compared with the other situations considered.  Most notably, already at around
$\sim40\,$ps after the pulse maximum it has dropped close to zero.  In sharp
contrast to the common interpretation that a Mandel parameter near zero implies
a distribution with a shape close to a Poissonian, Fig.~\ref{fig:chirp} (b)
shows a jagged distribution with two maxima at $\sim40\,$ps after the pulse
maximum.  Therefore, using the Mandel parameter as a measure for the deviation
from a Poisonian is not valid in all physically relevant situations.

\begin{figure*}[t]
	\centering
	\includegraphics[width=0.9\textwidth]{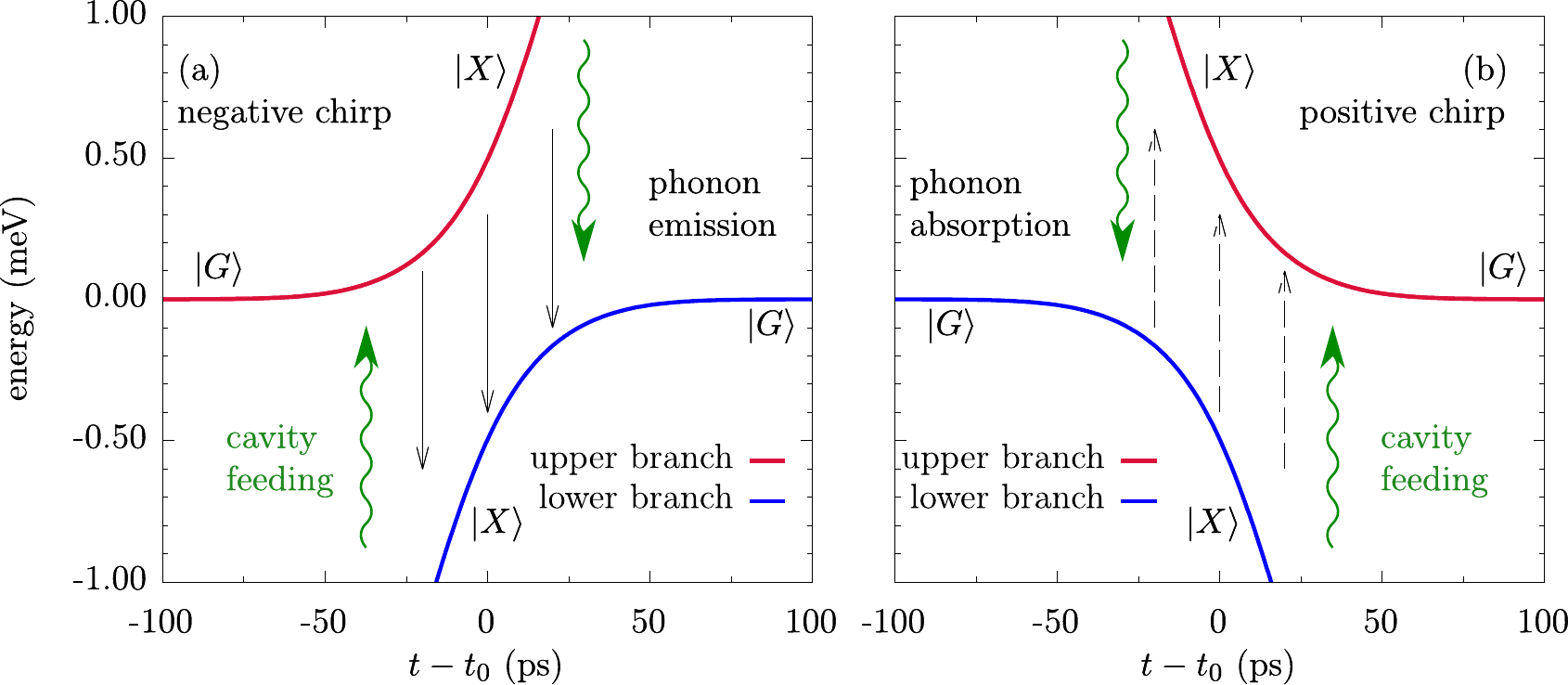}
	\caption{Time evolution of the upper and lower laser-dressed state
energies with respect to the excitation pulse maximum at $t=t_0$. While for
negative chirps (a) phonon emission is probable (represented by black arrows),
for positive chirps (b) phonon absorption is suppressed at low temperatures,
which is indicated by the dashed arrows. Green curly arrows indicate transitions
between laser-dressed states due to the QD cavity feeding.}
	\label{fig:dressed_states}
\end{figure*}

We further note that the Mandel parameter calculated for all excitation conditions
studied in this paper changes its sign during the course of time. 
Without chirp and low intensities (orange curve) this happens near the first maxima of
the $Q$ oscillations, as discussed above, but also for higher driving strength (blue curve)
a sign change occurs indicating that  before the pulse maximum is reached
the photon distribution is sub-Poissonian and switches at the pulse maximum to super-Poissonian.
Also for the chirped excitations $Q$ exhibits sign changes as revealed by the inset
in Fig.~\ref{fig:n_Q_fig_dyn} (b).
Actually, the Mandel parameter calculated for high pulse areas 
falls below zero before approaching its asymptotic
value of zero from below for chirped as well as for unchirped excitations.
Indeed, also the blue curve in Fig.~\ref{fig:n_Q_fig_dyn} (b) falls below zero
at $t-t_{0}=1090$ ps (not seen in the plotted range).
This sign change of $Q$ shortly before cavity losses have relaxed the photon distribution
to the empty cavity, can be understood as follows.
The maximal photon numbers that are transiently reached for high pulse areas are
well above one.  The cavity losses remove photons from the cavity such that
eventually the limit of $n=0$ with zero fluctuations is reached. However, since
the cavity losses for a state with $n$ photons scale like $\sim\,n$, the
relaxation from states with $n>1$ to lower states is faster than the final
relaxation from the $n=1$ to the $n=0$ states.  Therefore, before the final
relaxation is completed the photons preferably occupy the $n=1$ state which
results in a negative Mandel parameter before the asymptotic value of zero is
reached.

Finally, we note that $Q$ exhibits small amplitude oscillations for
chirped pulses which are absent in the unchirped case. A similar but less
pronounced tendency is seen in the mean photon number.

\subsection{Interpretation in terms of laser-dressed states}

A popular application of driving QDs with chirped laser pulses is the
robust preparation of exciton or biexciton states by invoking an adiabatic rapid
passage (ARP) process
\cite{Schmidgall2010,Simon2011,Wu2011,Lueker2012,Gawarecki2012,Debnath2012,Debnath2013,Glaessl2013,Mathew2014,Reiter2014,Kaldewey2017,PhysRevB.90.035316}.
ARP exploits the adiabatic theorem of quantum mechanics which predicts a time
evolution through instantaneous eigenstates (dressed states) of the system
provided the external driving fulfills the restrictions of the adiabatic regime
\cite{Born1928}.  In order to comply with these restrictions for a two-level
system driven by Gaussian chirped pulses with a frequency modulation given by
Eq.~(\ref{phi-chirp}), it is advisable to transform the QD-laser Hamiltonian
$H_{\t{DL}}$ in Eq.~(\ref{eq:H_DL_rotating}) to a frame co-rotating with the
phase $\varphi$ to get rid of a possibly rapidly changing coupling. The
transformed Hamiltonian reads:
\begin{align}
\label{eq:H_DL_rotating}
  \widetilde{H}_{\t{DL}}=&-\hbar\big(\D\w_{\t{LX}}+a\,(t-t_{0})\big)\XX\notag\\
  &-\frac{\hbar}{2}f(t)\left(\XG+\GX\right)\, .
\end{align}
The laser-dressed states can now be defined as the instantaneous
eigenstates of $\widetilde{H}_{\t{DL}}$. The corresponding eigenenergies are
plotted in Fig.~\ref{fig:dressed_states}, where the left panel corresponds to a
negative chirp while the result for positive chirp is shown in the right panel.
The distinctive feature of ARP is that when the system is in the ground state
$|G\rangle$ long before the pulse (i.e. for $t\to-\infty$) it will evolve
adiabatically towards the exciton state $|X\rangle$ after the pulse (i.e. for
$t\to+\infty$) independent of the sign of the chirp. However, it is important to
note that the evolution proceeds along the lower (upper) branch for positive
(negative) chirp. This affects in particular the impact of phonons. In general
phonons can efficiently induce transitions between the two branches. However, at
low temperatures phonon absorption is strongly suppressed and phonon emission
can invoke only transitions from the upper to the lower branch (cf.~the black
arrows in Fig.~\ref{fig:dressed_states}).  That is why phonons have little
effects on the ARP dynamics for positive chirp while for negative chirp the
ARP-based exciton preparation is strongly disturbed
\cite{Lueker2012,Debnath2012,Mathew2014,Reiter2014}.
In order to preserve an efficient exciton preparation also at negative chirps, it has been recently demonstrated that high pulse areas can be used since this effectively decouples the phonons from the electronic system \cite{Kaldewey2017,Reiter2019}.

When also a cavity is coupled to the QD, then the coupling leads to Rabi-type
rotations between states $|X,n\rangle$ and $|G,n+1\rangle$ with different numbers
$n$ of cavity photons.  In particular for times when the laser is far
off-resonant and the laser-dressed states are close to the undressed states, the
effect of coupling the QD to a cavity can be understood as inducing a transition
between the dressed states similar to the coupling to phonons.  To be a bit more
specific, when the system is in the exciton state the QD-cavity coupling leads
to a feeding of the cavity by an additional photon accompanied by a transition
from the $|X\rangle$-like branch to the $|G\rangle$-like branch (cf.~the green
curly arrows in Fig.~\ref{fig:dressed_states}).  At early times, the reverse process,
where one photon disappears from the cavity while transferring the system from
the ground to the exciton state is suppressed since there are initially no
photons in the cavity.
%We will see that this simple picture where the role of
%the cavity is reduced to cavity feeding accompanied by transitions
%$|X\rangle\to|G\rangle$ already explains a number of dynamical features found in
%our simulations.

We shall now try to interpret the pertinent features of the photon dynamics in
some more detail using the simplified picture where the system evolves
adiabatically trough the laser-dressed states in Fig.~\ref{fig:dressed_states}
while phonons and cavity feeding induce transitions between these states.

In the case of a negative chirp [cf.~Fig.~\ref{fig:dressed_states}(a)]
transitions form the upper branch to the lower branch of the laser-dressed
states accompanied by phonon emission are possible before and after the pulse
maximum at $t=t_0$. Thus, phonons should have a profound impact on the resulting
photon statistics during the entire pulse. In fact, this explains why
the distribution is
close to a thermal one at all times [cf.~Fig.~\ref{fig:chirp}(c)].
For times before the pulse reaches its maximum,
cavity feeding can occur form the exciton-like lower
branch to the upper branch, which has a large ground state contribution. 
Subsequently, the system can again decay to the lower branch by phonon
emission followed by another cavity feeding process back into the upper branch
and so on. Because of this constructive interplay between phonon and cavity
feeding processes, higher photon states can be reached compared with the
phonon-free situation for $t\leq t_0$
[cf.~Figs.~\ref{fig:chirp}(a) and (c)]. In the time
interval shortly after the pulse maximum the upper branch becomes the state with
the exciton-like characteristics and cavity feeding now takes place from the
upper branch into the ground state-like lower branch of the laser-dressed
states. Thus, after the pulse maximum has appeared phonon and cavity feeding
processes are now in direct competition with each other. Therefore,
compared with the phonon-free situation, the mean photon number should be reduced.
Altogether, for negative chirp, the phonon impact on the photon
distributions is visible at all times leading to nearly thermal distributions.
At times before the pulse maximum the interaction with
phonons increases the mean photon number because of a constructive interplay
between phonon and cavity feeding processes. This effect is reversed after the
pulse maximum and the mean photon number is reduced compared with the
phonon-free situation due to the phonon interaction, as can be seen
comparing the red with the gray curve in Fig.~\ref{fig:n_Q_fig_dyn} (a).\\

\begin{figure*}[t] \centering
\includegraphics[width=0.75\textwidth]{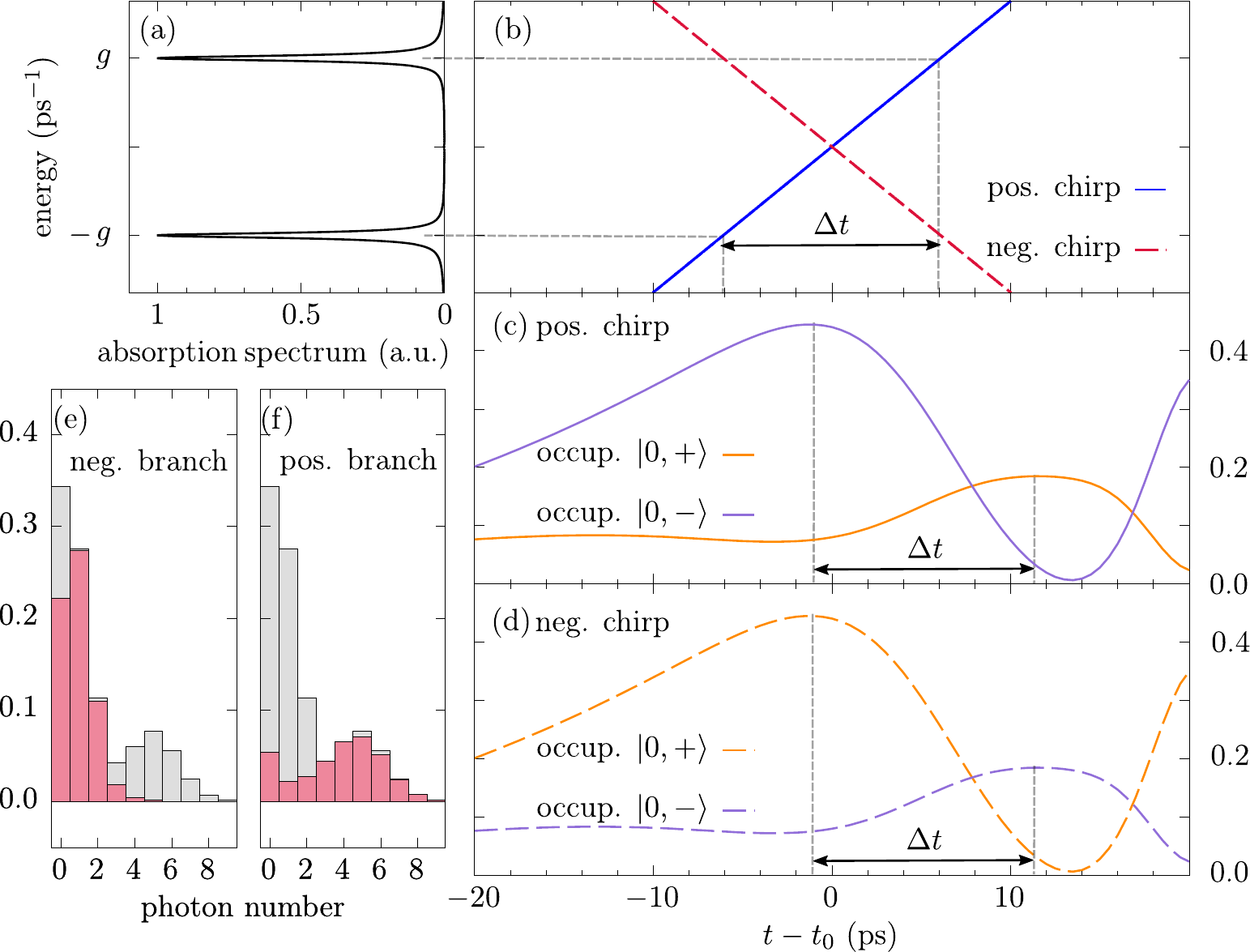}
\caption{(a) linear absorption spectrum of the QDC system. (b) time-dependent
  instantaneous frequency: blue (red) for positive (negative) chirp. $\Delta t$ marks
  the time elapsed between the crossing of the two resonances. (c) and (d) time evolution
  of the occupations of the lowest excited eigenstates of the QDC system [(c) for positive and
  (d) for negative chirp]. (e) and (f) photon number distribution at $t-t_{0}=20$ ps (gray),
  red: accounting only for $|n,+\rangle$ (e) or $|n,-\rangle$ (f) states.
  Here, only phonon-free results  are shown.}
	\label{fig:spectrum}
\end{figure*}

The situation is different when the chirp is positive as seen in
Figs.~\ref{fig:chirp}(b) and (d). Here, a phonon influence on the photon
statistics can be hardly seen before the pulse maximum.  This can again be
explained by inspection of the branches of the laser-dressed states.  Starting
in the ground state the system evolves adiabatically alongside the lower
branch. Since phonon absorption processes are suppressed at low temperatures,
transitions to the exciton-like upper state are unlikely to occur. Also cavity
feeding is hardly possible [cf. Fig.~\ref{fig:dressed_states}(b)] and, like in
the phonon-free situation, the system remains essentially in the ground state
without photons and phonons have almost no visible effect.
This observation changes after the pulse maximum.
Now, cavity feeding processes accompanied by transitions from the exciton-like lower
branch to the upper branch appear. Subsequently, phonon emission processes take
place, resulting in a transition back to the lower branch. Thus, now, a
constructive interplay between phonon emission and cavity feeding is possible,
leading to a thermalization of the photon distribution. Therefore, after a
transition time of a few 10\,ps the distribution resembles a thermal
distribution. Because of the constructive interplay the mean photon number is
increased compared with the phonon-free situation, as can be seen comparing
the cyan with the gray curve in Fig.~\ref{fig:n_Q_fig_dyn} (a).
Consequently, only for a finite time interval
after the pulse maximum photon distributions can be detected which are similar
to the distributions in the phonon-free situation and display irregular behavior
or several maxima.\\

\subsection{Interpretation in terms of cavity-dressed states}

Finally, we would like to explain why chirped pulse excitation leads to photon
number distributions where the number of maxima changes dynamically from one to
two and back to one. To this end we have to go beyond the laser-dressed states picture and
recall that the linear absorption of a QDC comprises two lines split by
$\Delta\omega=2g$ [cf.~Fig.~\ref{fig:spectrum}(a)].
Thus, the instantaneous frequency of a pulse with positive chirp
first crosses the energetically lower resonance and then, delayed by a time
$\Delta t= 2g/a$, the higher one [cf.~Fig.~\ref{fig:spectrum}(b)].
Each crossing of these resonances
initiates a wave-packet climbing up the Jaynes-Cummings ladder.
This behavior is efficiently described in the picture of the cavity-dressed states, i.e.,
the eigenstates of the dot-cavity Hamiltonian, which relate to the bare QD states by
\begin{align}
|n,+\rangle=\frac{1}{\sqrt{2}}\left(+|X,n\rangle + |G,n+1\rangle\right)\nn
|n,-\rangle=\frac{1}{\sqrt{2}}\left(-|X,n\rangle + |G,n+1\rangle\right)
\end{align}
in the case of a resonant cavity mode $\w_{\t{X}}-\w_{\t{C}}=0$.

Starting from the state $|G,0\rangle$ only the two states $|0,\pm\rangle$ can be reached directly
  by the laser coupling and thus climbing up the Jaynes cummings ladder one has to
  pass these states. Since the corresponding eigenenergies are separated
  by $2g$, the trasitions to these states are in resonance with the instantaneous
  frequency of a chirped pulse at different times.
Indeed, Fig.~\ref{fig:spectrum}(c) reveals that the occupation of the 
lowest excited eigenstate of the QDC system $|0,-\rangle$ rises
before the upper state $|0,+\rangle$ acquires a noticeable occupation.
The maximum occupation of $|0,-\rangle$ is reached $\approx\,5$ ps
after the instantaneous frequency has crossed the lower resonance,  revealing the
reaction time of the system. $|0,+\rangle$ is maximally occupied
delayed exactly by $\Delta t$ from the maximal occupation of $|0,-\rangle$.
The time ordering of the excitation of the $|0,\pm\rangle$ states is reversed when
reversing the sign of the chirp [cf.~Fig.~\ref{fig:spectrum}(d)] since now the
upper resonance is crossed first.

 The laser driving couples $|n,+\rangle$ to $|n,-\rangle$ states. However, when
  the instantaneous frequency is in resonance with transitions between $|n,+\rangle$ states with
  adjecent $n$ then the transitions to $|n,-\rangle$ states are off-resonant and vice versa.
  Thus, it can be expected that the packets running up the Jaynes-Cummings ladder
  are essentially composed either of $|n,+\rangle$ or $|n,-\rangle$ states. Indeed, this is
  confirmed by Fig.~\ref{fig:spectrum} (e) and (f) which displays in gray the photon number distribution
at time $t-t_{0}=20$ ps, i.e., the time where according to Fig.~\ref{fig:chirp} (b)
the two maxima are most pronounced. Also shown in red
are photon number distributions
calculated according to
\begin{align}
  P_{n}^{(\pm)} =\left\{
  \begin{matrix}
    &\frac{1}{2}\,(\langle n,\pm|\rho|n,\pm\rangle + \langle n-1,\pm|\rho|n-1,\pm\rangle)\\
     &\text{for } n>0,\\[1ex]
    &\frac{1}{2}\,(\langle 0,\pm|\rho|0,\pm\rangle + \langle G,0|\rho|G,0\rangle)\\
     &\text{for } n=0.
  \end{matrix}
       \right.
       \label{P+-}
\end{align} 
Recalling that for a cavity in resonance with the QD transition
the $|n,\pm\rangle$ states have a probability of $1/2$ for finding
$n$ or $n+1$ photons, Eq.~(\ref{P+-}) yields, for $n>0$, the
probability for having $n$ photons when accounting only for
either the $|n,+\rangle$ or the $|n,-\rangle$ states.
For $n=0$ the contribution from $|G,0\rangle$ is counted
by $1/2$ for the plus and minus branch, since this state
can be counted as lower or upper state.
We note in passing that  $P_{n}^{(-)}$  [red bars in
Fig.~\ref{fig:spectrum} (e)] does not add up with
$P_{n}^{(+)}$  [red bars in Fig.~\ref{fig:spectrum} (f)]
to the total photon number $P_{n}$ [gray bars in Fig.~\ref{fig:spectrum}],
because $P_{n}$ comprises coherences between the $|n,+\rangle$ and the $|n,-\rangle$ states
in addition to their occupations. Nevertheless,  Fig.~\ref{fig:spectrum} reveals that
the two peaks in the photon number distribution can be attributed
unambiguously either to the upper or lower branch of the QDC states.

Altogether this explains the time evolution of the peaks in the photon number distribution.
After crossing the first resonance the distribution has a single peak since at first
only a single packet is climbing up the Jaynes-Cumming ladder. When the second
resonance is crossed a second packet is initiated such that at $t-t_{0}\approx 20$ ps
two well resolved packets are observed. Both packets move up and down
the Jaynes-Cummings ladder similar to the single wave-packet observed for
the unchirped excitation in Fig.~\ref{fig:chirp0} (b), (d). Since the decline
of the first packet starts while the second is still rising, at some time
both packets overlap. Although the packets are no longer well resolved, two maxima
are still found over an extended time period
[$30\,\text{ps}\lesssim t-t_{0}\lesssim 50$ ps in Fig.~\ref{fig:chirp} (b)].
At later times the relaxation drives both packets to low photon numbers
such that the maxima merge and a single-peaked distribution is recovered.

Finally, we note that for a cavity in resonance with the QD transition the
energies of the QDC eigenstates $|n,\pm\rangle$ are found in the rotating frame
at $\hbar\omega_{n,\pm}=\pm g\,\sqrt{n+1}$ such that the transition energies
between states with adjacent $n$ are all different and decrease with rising $n$.
Therefore, the instantaneous frequency of a chirped pulse crosses all of these
resonances at different times which is likely to contribute to
the somewhat irregular looking time evolution of the photon number distribution
found in particular in the intermediate time interval
$30\,\text{ps}\lesssim t-t_{0}\lesssim 50$ ps in Fig.~\ref{fig:chirp} (b).

\section{Conclusion}
\label{sec:Conclusion}

We have studied transient photon number distributions generated in
a microcavity by a pulsed excitation of an embedded quantum dot.
We find qualitatively different photon distributions for chirped
and unchirped pulses. Phonons have a noticeable influence on the
photon distributions in particular for negative chirps, where
the phonon coupling introduces qualitative changes of the shape of the
distribution already at a temperature of $T=4$ K. To be more specific,
phonons lead in this case to almost thermalized photon distributions
at high effective temperatures for all times. For positive chirp, the transient
distributions are far away from a thermal one for times after the pulse maximum until
about 80 ps afterwards.

For all investigated cases, we find that the Mandel parameter changes its sign
during the time-evolution of the system, indicating the ability to enter and
leave a regime of genuine non-classical photon statistics in the course of time.
Moreover, cases were encountered where the Mandel parameter is zero, but the
photon number distribution has two peaks and is definitely not a Poissonian.
Therefore, one has to be careful when using the Mandel parameter as a measure
for the deviation from a Poissonian distribution, as it is often done.  This
finding underlines the necessity to carefully consider the definition of the
Mandel parameter, which indeed yields zero for a Poissonian distribution. But
the reverse implication is obviously not true for all cases.

Our most striking result is, however, that the shape of the photon number
distribution changes significantly during the time evolution when the system is
excited by chirped pulses.  In fact, when the excitation starts to populate
states with higher photon numbers, one observes at first bell-shaped
distributions with a single maximum that increases in time.  Subsequently, two
well separated bell-shaped contributions develop which at later times first
evolve into a single broad feature with two peaks and eventually merge into a
distribution with a single peak.  This is in sharp contrast to the unchirped
case, where for the same high driving strengths the photon number distributions
keep a bell-shape with a single maximum for all times. Our analysis reveals
that the transient changes of the shape of the photon distribution in the
chirped case can be attributed to subsequent crossings of resonances of the
quantum-dot--cavity system by the instantaneous frequency.

We believe that our findings deepen the understanding of the transient
behavior of photon distributions in a driven quantum-dot--cavity system
and its dependence on the driving conditions. This might  pave the way to targeted
manipulations of photon distributions which could result in new types of
photonic applications in the future.

\acknowledgments
M.Cy. thanks the Alexander-von-Humboldt foundation for support
through a Feodor Lynen fellowship.  A.V. acknowledges the support from the
Russian Science Foundation under the Project 18-12-00429 which was used to study
dynamical processes non-local in time by the path-integral approach.  This work
was also funded by the Deutsche Forschungsgemeinschaft (DFG, German Research
Foundation) - project Nr. 419036043.

\bibliography{bib}
\end{document}